\documentclass[prl,twocolumn,amsmath,amssymb]{revtex4}
\usepackage{graphicx}
\usepackage{hyperref}
\usepackage{color}

\usepackage{epsf}
\usepackage{amssymb}

\newcommand{\la}{\label}
\newcommand{\be}{\begin{equation}}
\newcommand{\ee}{\end{equation}}
\def\12{\frac{1}{2}}

\begin{document}
\title{Harmonic measure of critical curves}
\author{E. Bettelheim}
\author{I. Rushkin}
\author{I. A. Gruzberg}
\author{P. Wiegmann}
\altaffiliation[Also at ]{Landau Institute of Theoretical
Physics.}
\affiliation{James Frank Institute, University of
Chicago, 5640 S. Ellis Ave. Chicago IL 60637}
\date{\today}

\begin{abstract}
Fractal geometry of critical curves appearing in 2D critical
systems is characterized by their harmonic measure. For systems
described by conformal field theories with central charge
$c\leqslant 1$, scaling exponents of the harmonic measure have
been computed by B. Duplantier [Phys. Rev. Lett. {\bf 84}, 1363
(2000)] by relating the problem to boundary two-dimensional
gravity. We present a simple argument connecting the harmonic
measure of critical curves to operators obtained by fusion of
primary fields, and compute  characteristics of the fractal
geometry by means of regular methods of conformal field theory.
The method is not limited to theories with $c\leqslant 1$.
\end{abstract}
\maketitle

\paragraph{Introduction.}

Two-dimensional (2D) statistical models typically exhibit stochastic
curves, such as external perimeters of critical clusters in the
Potts model or fluctuating loops in models of Refs.
[\onlinecite{nienhuis}--\onlinecite{kondevhenley}]. In the critical
regime these curves are fractal. Different phases of loop models
exhibit two distinct types of critical curves: dilute and dense.
Dilute curves are simple while dense curves have infinitely many
double points at which they touch but do not intersect. There exists
a duality relation between the two types, in particular, the
external perimeter of a dense curve is a dilute curve
\cite{duplantier}. Despite of a long history of studying critical
phenomena, inquiries into the stochastic geometry of the critical
curves are relatively recent. Traditional conformal field theory
(CFT) \cite{cft} concentrates on correlations of local fields and
essentially leaves their relation to geometrical objects, such as
critical curves, unclear. B. Duplantier, in a seminal paper
\cite{duplantier}, found the harmonic multifractal spectrum of
critical curves arising in critical systems with the central charge
$c\leqslant 1$ (see Eqs. (\ref{longresults}, \ref{9})). His method
used an intriguing connection between CFT and 2D boundary quantum
gravity. More recently, the stochastic Loewner evolution approach
was successfully used for the same purpose \cite{slepeople}. Both
methods are interesting and powerful, but somewhat foreign to
traditional CFT approach  and not obviously susceptible to
generalizations.

In this letter we show that geometrical properties of dilute
critical curves are naturally linked to correlation functions of
primary fields and can be studied within the regular framework of
CFT, not limited to theories with  $c\leqslant 1$. Our  arguments
may also shed some light on the boundary quantum gravity itself.

\paragraph{Harmonic measure of critical curves.}

A basic characteristic  of simple curves is the harmonic measure
\cite{ahlfors}. It has a simple electrostatic interpretation.
Consider a closed  curve $\gamma$ made of a conducting material
and carrying a total electric charge of one. Harmonic measure of
any part of  the curve is the charge of this part. In what follows
we will pick a point of interest $z_0$ on the curve and consider a
disc of a small radius $r$ centered at $z_0$ as is in the Figure.
It surrounds a small part of the curve, and we define $\mu(r)$ to
be harmonic measure of this part.

If  a curve  is a closed loop lying entirely in the bulk (i.e. it
does not touch the system boundaries), all its points are
statistically identical. We can also consider the case when a
curve emerges from a boundary, and the case when a curve has one
or both dangling ends in the bulk. The latter is achieved by
inserting into the system a small isolated boundary (almost a
puncture) so that the curve ends on this boundary. Statistics of
harmonic measure is different in the following three cases
\cite{duplantier} depicted in the Figure: (i) {\it bulk}, if $z_0$
lies in the bulk and is not an endpoint; (ii) {\it boundary}, if
$z_0$ is a point where $\gamma$ is connected with the system
boundary; (iii) {\it extremity}, if $z_0$ is a dangling endpoint
of $\gamma$ in the bulk. More generally one may consider $n$
curves emanating from the boundary, or $n$ curves meeting at a
point in the bulk. The bulk and extremity cases correspond to
$n=2$ and $n=1$.
\begin{figure}[h111]
\begin{center}
\includegraphics[width=2.5in]{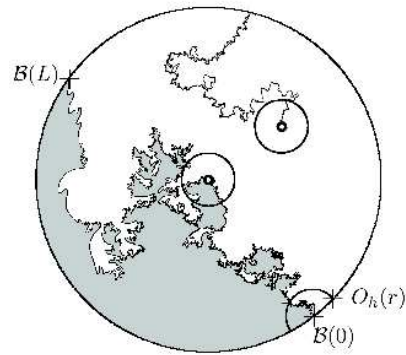}
\caption{
Harmonic measure of a critical curve evaluated  in the
bulk, on the system  boundary, and at an extremity.}
\end{center}
\end{figure}

Let us define a conformal map $w(z)$ of  the exterior of $\gamma$ to
some standard domain. For a closed loop in the bulk it can be the
exterior of a unit disc, while for a curve with both ends on the
system boundary or a ``sprout", the upper half plane is more
convenient. We normalize the map so that the point of interest $z_0$
is mapped onto itself, choose it to be the origin $z_0=0$, $w(0)=0$,
and demand that $ w'(\infty)=1$. The moments of harmonic measure of
dilute critical curves scale at small $r$ with non-trivial
exponents. The relations $\mu(r)\sim |w(r)|\sim r |w'(r)|$ follow
from the definition of $\mu(r)$. Therefore, the scaling of the
moments of $\mu(r)$ is determined by that of $|w'(r)|$:
\begin{align}\label{electricfield}
\prec\!\! |w'(r)|^h\!\!\succ \; \sim \; r^{\Delta(h)},\quad
r\rightarrow 0.
\end{align}
The notation $\prec\!\!\dots\!\!\succ$ stands for statistical
average over the ensemble of critical curves. The notation
$\langle\dots\rangle$ is reserved for correlators of CFT.

\paragraph{Harmonic measure and fluctuating geometry.}

The scaling of the harmonic measure may be computed by considering
CFT correlation functions. It is the easiest to start with a point
where a curve $\gamma$ connects with the system boundary. We
assume that the system occupies the interior of an annulus. The
curve starts from  the large circle at $z_0=0$, ending on the
large circle at a point $L$.

The partition function $Z(0,L)$ restricted to configurations that
contain a curve $\gamma$ connecting the points 0 and $L$, is given
by the correlator of two appropriate boundary operators
$\mathcal{B }$ changing boundary conditions \cite{cardyoperators}.
These operators create curves emanating from the boundary
\cite{duplantiersaleur}:
\begin{align}
Z(0,L)/Z = \langle \mathcal{B }(0)\mathcal{B
}(L)\rangle_\mathbb{H},
\end{align}
where $Z$ is the unrestricted partition function. The subscript of
$\langle\dots\rangle$ refers to the domain of definition, in this
case the upper half plane.

This correlator, as indeed any correlator containing two boundary
operators $\mathcal{B}$,  can be computed in two steps: In the
first step  we pick a particular realization of the curve
$\gamma$. Within each realization, the curve $\gamma$ is the
boundary separating two independent systems---the interior and the
exterior of $\gamma$---with partition functions
$Z^\mathrm{int}_\gamma$ and $Z^\mathrm{ext}_\gamma$, respectively.
These are stochastic objects that depend on the {\it fluctuating
geometry} of $\gamma$. At the second step we sum over the
realizations of $\gamma$. We thus obtain
\begin{align}
Z(0,L) = \prec\!\! Z^\mathrm{int}_\gamma
Z^\mathrm{ext}_\gamma\!\!\succ.
\end{align}

We further insert an additional boundary primary operator $O_h(r)$
of conformal weight $h$ sufficiently close to 0, and a second copy
$O_h(\infty)$. Both act as a source of harmonic measure. We thus
consider the correlator
\begin{align}\label{thecorr}
\langle O_h(r) \mathcal{B}(0) \mathcal{B}(L) O_h(\infty)
\rangle_\mathbb{H}.
\end{align}
Since we are only interested in the $r$-dependence of the
correlator, we can fuse together the distant primary fields:
$\mathcal{B }(L)\times O_h(\infty)=\Psi(\infty)$. We therefore
consider the $r$-dependence of a 3-point function
\begin{align}\label{the3corr}
\langle O_h(r) \mathcal{B}(0) \Psi(\infty) \rangle_\mathbb{H},
\end{align}
and show that it yields the statistics of the harmonic measure.

Decomposing the upper half plane into the exterior and the
interior of $\gamma$ as before, we can rewrite (\ref{thecorr}) as
the average over the fluctuating geometry of $\gamma$:
\begin{align}\label{7}
\prec\!\! \langle O_h(r) O_h(\infty) \rangle^\mathrm{ext}_\gamma
Z^\mathrm{int}_\gamma Z^\mathrm{ext}_\gamma\!\!\succ/ Z
\end{align}
Here the domain of the definition of the  correlator of primary
fields is the exterior of $\gamma$. This correlator is
statistically independent from the other two factors  in the
numerator of (\ref{7}) in the limit $r\ll |L|$. We are left with
the correlation function $\langle
O_h(r)O_h(\infty)\rangle^\mathrm{ext}_\gamma$ of two primary
fields of boundary CFT, further averaged over all configuration of
the boundary $\gamma$. It is equal to  the 3-point correlation
function (\ref{the3corr}).

Now we  apply  the conformal transformation $w(z)$ which maps the
exterior of $\gamma$ onto the upper half plane. Being a primary
operator of conformal weight $h$,  $O_h(r)$ transforms as
$O_h\rightarrow O_h(w(r))|w'(r)|^h$, while $O_h(\infty)$ does not
change because of the normalization of $w(z)$. The transformation
relates the correlation function in the exterior of $\gamma$ to a
correlation function in the upper half plane: $\langle
O_h(w(r))O_h(\infty)\rangle_\gamma^\mathrm{ext}=|w'(r)|^h \langle
O_h(w(r))O_h(\infty)\rangle_\mathbb{H }$. The latter does not
depend on $r$, but rather on the size of the entire system.

Summing up, we obtain a scaling relation between the moments of
the harmonic measure near the boundary and correlation functions
of primary operators  \cite{bernard}:
\begin{align}\la{goodrelation}
\langle O_h(r) \mathcal{B}(0) \Psi(\infty)\rangle_\mathbb{H}
\sim\;\prec\!\! |w'(r)|^h\!\!\succ,\quad r\ll |L|.
\end{align}
This scaling relation is the main result of the paper. It allows
us to reproduce the scaling of the harmonic measure upon
identification of the operators $\mathcal{B}$.

\paragraph{The spectrum of the harmonic measure for CFTs
with $c \leqslant 1$.}

The scaling exponents of $\prec\!\! |w'(r)|^h\!\!\succ$ in the bulk,
near the boundary and near an extremity are denoted
$\Delta_{\rm{bulk}}(h),$ $\Delta(h)$ and $\Delta_{\rm{extr}}(h)$,
respectively. If we parametrize the central charge of the critical
system by $c = 1-6\big(\sqrt{4/\kappa}-\sqrt{\kappa/4}\big)^2$ where
$0 < \kappa\leqslant 4$ for dilute and $\kappa>4$ for dense critical
curves, then the exponents obtained in \cite{duplantier} read
\begin{align}
\Delta_{\rm{extr}}&= -h/2 +\Delta \kappa/8 , \quad
\Delta_{\rm{bulk}}= \Delta_{\rm{extr}} + \Delta/2  \label{9}
\end{align}
The boundary exponent happens to be identical to the dressed
gravitational dimension. This is the dimension of a primary field
on a random surface, which dimension on a flat space is $h$. They
are connected by equation \cite{KPZ}
\begin{align}\label{longresults}
\frac{\Delta(\Delta-\gamma_{\rm{string}})}{1-\gamma_{\rm{string}}}=h,
\quad 1-\gamma_{\rm{string}}=\frac{4}{\kappa}.
\end{align}

In order to derive these results  along the lines presented above,
we must identify the operators creating critical curves.  In what
follows this  result  and its generalizations to the case of $n$
curves meeting at the same point, will appear  in a compact form
in Eqs.  (\ref{eb}, \ref{taubulk}) in terms of the charges of
relevant  operators.

\paragraph{Levels of a Gaussian field.}

The relation between critical curves and boundary CFT is most
transparent in the case when they are dense ($\kappa>4$) upon
representation as level lines of a Gaussian field
\cite{nienhuis,cft,kawai}. Below we discuss this representation and
find the operators representing dense critical curves. The duality
then implies that the corresponding operators for the dilute curves
are obtained by simple continuation to $\kappa\leqslant 4$.

A $c\leqslant 1$ CFT in an annulus $D$ is described by a real
compactified Bose field $\varphi(z,\bar z)$. In the normalization
where the radius of compactification is equal to $1$, so that $
\varphi\simeq\varphi+2\pi,$ the classical action reads
\cite{kondev}:
\begin{align}\la{C}
S = \frac{g}{4\pi }\int_D |\nabla \varphi|^2 +
i\frac{e_0}{2\pi}\int_{\partial D} K\varphi+\int_D e^{2i\varphi}.
\end{align}
Here $K$ is the geodesic curvature of each component of the system
boundary and the stiffness $g$ and the ``charge" $e_0$ are related
to $\kappa$ as $g=4/\kappa,\,e_0=1-4/\kappa$. The last term in the
action (\ref{C}) represents the marginal part of a $2\pi$-periodic
locking potential.

We impose the boundary condition
\begin{align}\la{D}
\partial_n\varphi |_{\partial D} = 0,
\end{align}
where the derivative is normal to the boundary. The real field
$\varphi$ is the sum of a holomorphic and an antiholomorphic parts
and a real zero-mode $\phi_0$:
\begin{align}
\varphi(z,\bar z) = \phi(z) + \overline{\phi(z)} +
\phi_0.
\end{align}
The fields $\phi(z)$ and $\overline{\phi(z)}$ are glued together on
the boundary by the condition (\ref{D}) which reads $\partial_l
\phi(z) =
\partial_l \overline{\phi(z)}$, where $\partial_l$ is tangential derivative. Both parts can be considered as
one holomorphic field on a torus (Schottky double) obtained by
gluing the annulus to its reflected copy along the boundaries. The
``dual'' Bose field $\tilde\varphi(z,\bar z)=
(4/i\kappa)(\phi(z)-\overline{\phi(z)})$ is related to $\varphi$
through Cauchy-Riemann conditions.

The gradient $ J = \nabla \varphi(z,\bar z)$ is therefore the
current. It is conserved. On the boundary $|z|=1$ the boundary
condition (\ref{D}) then means that no vector current flows through
the system boundary $J_n|_{\partial D}=0$.

\paragraph{Fluctuating loops.}

The critical curves have a simple interpretation in terms of the
Bose field. They are the level lines $\mathrm{Re}\,\varphi(z,\bar z)
=k\pi$ of the height function $\mathrm{Re}\,\varphi(z,\bar z)$. The
level lines are non-intersecting plane loops which are identified
with boundaries of critical clusters in the  Potts model, or lines
in the $O(n)$ model
[\onlinecite{nienhuis}--\onlinecite{kondevhenley}]. In this
formulation, statistical models represented by CFTs with $c\leqslant
1$ can be seen as a gas of fluctuating loops. The boundary condition
ensures that the loops cannot cross the system boundaries and,
furthermore, that the system boundaries themselves are loops.

In the Hamiltonian formalism of radial quantization, the
non-contractible loops $\cal C$ in the annulus represent  coherent
states propagating along the cylinder $\zeta = \log z$. Coherent
states are defined by the condition $\varphi(z,\bar z)|{\cal
C},\phi_0\rangle= \phi_0|{\cal C},\phi_0\rangle$, $z\in {\cal C}.$
Contractible loops represent virtual states. The partition
function with the action (\ref{C}) can be seen as the overlap of
the boundary coherent states.

A normalization of $\varphi$, customary in the CFT literature,
fixes $g=\frac{1}{2}$ and introduces the notation
$\alpha_+=\sqrt{4/\kappa},
\,\alpha_-=-\sqrt{\kappa/4},\,2\alpha_0=\alpha_++\alpha_-$
\cite{cft}. The radius of compactification is then ${\cal
R}=\sqrt{8/\kappa}=\sqrt{2}\alpha_+$. Below we proceed in the
physical normalization, where  ${\cal
R}=1$.

\paragraph{Primary operators.}

In terms of the Bose field  the primary operators read
\begin{align}\label{boundarybulk}
\mathcal{O}^{(e,m)}(z,\bar{z})= e^{ie\varphi(z,\bar{z})}
e^{m\tilde\varphi(z,\bar{z})},
\end{align}
where $e$ and $m$ are ``electric" and ``magnetic" charges.

The holomorphic weight of this operator is $h(\alpha) =
\alpha^2-2\alpha\alpha_0,$ where $ \alpha= -\frac{1}{2}(e\alpha_-
+ m\alpha_+)$ is the holomorphic charge. The antiholomorphic
charge of (\ref{boundarybulk}) is
$\bar\alpha=-\frac{1}{2}(e\alpha_- - m\alpha_+)$. It is also
customary to label the primary operators and their holomorphic
charges by two  numbers $r,s$ using the Kac table $
\alpha_{r,s}=\frac{1}{2}(1-r)\alpha_++\frac{1}{2}(1-s)\alpha_-.$
Here $r=1+m$, $s=1+e$.

The magnetic operator with $e=0,\, m=1$ introduces a defect line
on which the value of the field $\varphi$ changes by the
compactification length $2\pi$. The electric operator with $e=1,\,
m=0$ picks up a phase difference $2\pi$ while going around the
magnetic operator. A general $\mathcal{O }^{(e,m)}$ is the
composition of the two.

\paragraph{Operators representing critical curves
\cite{duplantier,cardyoperators,duplantiersaleur}.}

The magnetic operator with charge $m$, applied at the boundary,
changes the values of $\varphi$ by $\pi m$ (since one can go from
one side of the operator to the other by half of a full turn):
${\cal O}^{(0,m)}(0)\varphi(x)=(\varphi(x) + m \pi \theta(x)){\cal
O}^{(0,m)}(0)$, where $\theta(x)$ is the step function and $x$ is
the coordinate along the boundary. Thus, it is a boundary condition
changing operator. In particular, ${\cal O}^{(0,1)}(0)$ changes the
boundary condition by half the compactification length, just as a
cluster wall does, so it produces a critical curve emanating from
the system boundary at $x=0$ in the direction normal to the
boundary. This is, therefore, the previously mentioned operator
$\mathcal{B}$. It happens to be degenerate on level 2 and is placed
in  the Kac table as\footnote{It is also common in the literature to
exchange the definitions $\alpha_+\leftrightarrow-\alpha_-$. In the
Kac notation it means $r\leftrightarrow s$.} $\Psi_{2,1}$. We can
insert two such operators on the external boundary of the annulus
and the other two on the central puncture in order to connect the
boundaries with two curves. These curves then divide the annulus
into two domains, each having the topology of a disc.

Similarly, the boundary operator ${\cal O}^{(0,n)} \equiv
\Psi_{n+1,1}$ is degenerate on level $n + 1$ and produces $n$
curves. Its holomorphic charge is
$\alpha_{n+1,1}=-\frac{n}{2}\alpha_+$.

\paragraph{Pinning operator.}

This operator creates $n$ curves emanating from a point in the
bulk, that is, from a puncture. The puncture, as any internal
system boundary, carries the electric charge $-e_0$. The
combination of this charge and the magnetic charge associated with
the creation of ${n}$ curves identifies the bulk pinning operator
as ${\cal O}^{(-e_0,{n}/2)}$ with the conformal charge
$\alpha_{1+n/2,4/\kappa} \equiv \alpha_{\frac{n}{2},0}=
-\frac{{n}}{4}\alpha_++\alpha_0.$ In the Kac classification it is
the field $\Psi_{\frac{n}{2},0}$.

\paragraph{Screening operators.}

The bulk magnetic operator with the double magnetic charge ${\cal
O}^{(0,2)}$  cuts a loop and changes the direction of each part by
$\pm \pi$ giving rise to a new loop representing a virtual state.
This operator  must be marginal, which is the origin of the
relation between the stiffness and the screening charge $e_0+g=1$.
Another marginal (i.e. screening) operator ${\cal O}^{(2,0)}$
represents the marginal part of the locking potential in
(\ref{C}).

\paragraph{Boundary exponents.}

We argued that the curve-creating operator $\mathcal{B }$ in
(\ref{goodrelation}) is $\Psi_{2,1}$. Therefore,
(\ref{goodrelation}) reads
\begin{align}\la{O}
\langle O_h(r)\Psi_{2,1}(0)\Psi(\infty)\rangle_{\mathbb{H
}}\sim\;\prec\!\! |w'(r)|^h\!\!\succ,
\end{align}
where the holomorphic charge of $O_h$ is $\alpha_h = \alpha_0 -
\sqrt{\alpha_0^2 + h}$ and that of $\Psi(\infty)$ is $2\alpha_0 -
\alpha_h - \alpha_{2,1}$ (the choice of dual charges is made
unique by $\Delta(0)=0$.)

On the other hand, the scaling behavior of this correlator is
easily computed by regular means of Coulomb gas technique
\cite{cft}. It scales as $r^{2\alpha_h\alpha_{2,1}}$. The
comparison yields the result (\ref{longresults}) written in a
suggestive form:
\begin{align}\la{eb}
\Delta(h) = 2\alpha_h\alpha_{2,1}, && \alpha_h &= \alpha_0 -
\sqrt{\alpha_0^2 + h}
\end{align}
where string susceptibility
$\gamma_{\rm{string}}=1-4\alpha_{2,1}^2$.

An immediate generalization of this formula  $\Delta^{(n)} = 2
\alpha_h \alpha_{n + 1, 1} = n \Delta $  can be obtained by
replacing $\Psi_{2, 1}$ by $\Psi_{n + 1,1}$ in (\ref{O}). It
describes the harmonic measure of ${n}$ curves reaching the system
boundary at one point.

\paragraph{Bulk exponents.}

Similar arguments yield the value of the bulk exponents. Let
points $0$ and $L$ lie in the bulk and consider a bulk correlator
\begin{align} \la{bulkcorrelator} \langle
O_{h'}(r)\Psi_{1,0}(0)\Psi_{1,0}(L)
O_{h'}(\infty)\rangle.
\end{align}
The fields $\Psi_{1,0}$ ensure the existence of a closed curve
$\gamma$ connecting $0$ and $L$. In the limit $r\ll |L|$ and
because we are only interested in the $r$-behavior, we can fuse
$\Psi_{1,0}(L)\times O_{h'}(\infty)=\Psi(\infty)$ where the charge
of $\Psi$ is $2\alpha_0-\alpha_{h'}-\alpha_{1,0}$, and consider
instead a 3-point function $\langle
O_{h'}(r)\Psi_{1,0}(0)\Psi(\infty)\rangle$.

As in the boundary case, we argue that this correlator equals
$\prec\!\! \langle
O_{h'}(r)O_{h'}(\infty)\rangle_\gamma^\textrm{ext}\!\!\succ$ up to
a normalization. We transform it into a correlator in a disc
exterior via a map $w(z)$ which takes $\gamma$ to a unit circle
centered at $-i$ such that $O_{h'}(r) \to
|w'(r)|^{2h'}O_{h'}(w(r)).$ The difference with the boundary case
is that $O_{h'}$ is now a bulk field in the presence of a circular
 boundary. We therefore fuse $O_{h'}(w(r))$ with its image inside
the disc (approximately at $w^*(r)$) and take the leading
non-trivial fusion product $(w(r) - w^*(r))^{h-2h'}O_{h}(0)$. Here
$O_{h}$ is a primary field of weight $h$. We determine $h$ through
the neutrality condition $2\alpha_{h'} +(2\alpha_0 - \alpha_{h}) =
2\alpha_0$, yielding $\alpha_h = 2\alpha_{h'}$.

Since $r$ is small, $w(r) - w^*(r) \sim r|w'(r)|$. Summing up, we
give the scaling behavior of the original correlator:
\begin{align} \label{cftef}
\langle O_{h'}(r)\Psi_{1,0}(0)\Psi(\infty)\rangle\sim
r^{h-2h'}\prec\!\! |w'(r)|^{h}\!\!\succ.
\end{align}
On the other hand, scaling behavior of this correlator at small
$r$ is easily found by regular CFT means: it scales as
$r^{4\alpha_{h'}\alpha_{1,0}}$. We thus obtain
\begin{align} \la{taubulk}
\Delta_{\rm{bulk}}(h) = 2h' - h + 4 \alpha_{h'} \alpha_{1,0},
\end{align}
which coincides with the result given in Eq. (\ref{9}).

A replacement $\Psi_{1,0}\to\Psi_{\frac{n}{2},0}$  in
(\ref{bulkcorrelator}) produces $n$ curves emanating from a given
bulk point.  Their scaling exponents are obtained by replacing
$\alpha_{1,0}\to\alpha_{\frac{n}{2},0}$ in  (\ref{taubulk})
\begin{align} \la{taubulk1}
\Delta_{\rm{bulk}}^{(n)}(h) = -h/2 +
\alpha_{h}(2\alpha_{n/2,0}-\alpha_0).
\end{align}
If $n$ is even it gives a bulk exponent for the case when $n/2$
curves are passing through one point. The case $n=1$ yields the
extremity exponent $\Delta_{\rm{extr}}(h)$ of a single dangling
end in the bulk (\ref{9}).

\paragraph{Discussion}

The method of computing the statistics of the harmonic measure,
discussed here, is amenable to generalizations. For instance, one
can compute multipoint correlation functions. Another
generalization is for curves generated by conformal field theories
with $c>1$ \cite{kondev,kondevhenley}. In particular, it has been
shown in \cite{BGLW}  that SU$(2)_k$ - Wess-Zumino-Witten model
generates critical curves identical to those generated by its
coset $\text{su}(2)_k \oplus \text{su}(2)_1/\text{su}(2)_{k+1}$
--- a unitary minimal model with $\kappa= 4\frac{k + 2}{k + 3}$.
Its substitution into Eqs. (\ref{longresults}, \ref{9}) gives the
scaling of the harmonic measure for this case.

We benefitted from discussions with  P. Di Francesco, B.
Duplantier, L. Kadanoff, and I. Kostov. Our special thanks to P.
Oikonomou who made the picture and to N. Yufa for her help. This
work was supported in part by the NSF under  DMR-0220198 (PW). EB,
IG and PW were supported by NSF MRSEC Program under DMR-0213745.
IG  was supported by the Alfred P. Sloan Foundation and the
Research Corporation.

\end{document}